\documentclass{iaus}
\usepackage{graphicx}
\usepackage{natbib}

\newcommand{\sm}{M$_{\odot}$}

\newcommand{\kms}{\ensuremath{{\rm km~s}^{-1}}}
\newcommand{\urltilde}{\kern -.15em\lower .7ex\hbox{~}\kern .04em}

\def\aap{A\&A}
\def\apj{ApJ}

\def\aj{AJ}
\def\apjl{ApJL}
\def\mnras{MNRAS}

\def\nat{Nature}
\def\araa{ARA\&A}

\title[SNe Ib/c with and without GRBs] 
{Type Ib/c Supernovae with and without Gamma-Ray Bursts}

\author[Maryam Modjaz]   
{Maryam Modjaz$^1$}

\affiliation{$^1$CCPP, New York University, 4 Washington Place \\ New York City,
NY, 10003, USA \\ email: {\tt mmodjaz@nyu.edu} \\[\affilskip]
}

\pubyear{2012}
\volume{279}  
\pagerange{xxx--yyy}
\jname{Death of Massive Stars: Supernovae and Gamma-Ray Bursts}
\editors{P. Roming, N. Kawai \& E. Pian, eds.}
\begin{document}

\maketitle

\begin{abstract}
While the connection between Long Gamma-Ray Bursts (GRBs) and Type Ib/c Supernovae (SNe Ib/c)
from stripped stars has been well-established, one key outstanding question is what conditions and 
factors lead to each kind of explosion in massive stripped stars. One promising line of
attack is to investigate what sets apart SNe Ib/c \textbf{with} GRBs from
those \textbf{without} GRBs. Here, I briefly present two observational studies that probe the SN 
properties and the environmental metallicities of SNe Ib/c (specifically broad-lined SNe Ic) with and 
without GRBs. I present an analysis of expansion velocities based on published spectra 
and on the homogeneous spectroscopic CfA data set of over 70 SNe of Types IIb, Ib, Ic and Ic-bl, 
which triples the world supply of well-observed Stripped SNe. Moreover, I demonstrate that a 
meta-analysis of the three published SN Ib/c metallicity data sets, when including only values at 
the SN positions to probe natal oxygen abundances, indicates at very high significance that 
indeed SNe Ic erupt from more metal-rich environments than SNe Ib, while SNe Ic-bl with GRBs 
still prefer, on average, more metal-poor sites than those without GRBs.


\end{abstract}

\firstsection 
\section{Introduction}

\begin{figure*}[!ht]

\includegraphics[scale=0.37,angle=-90]{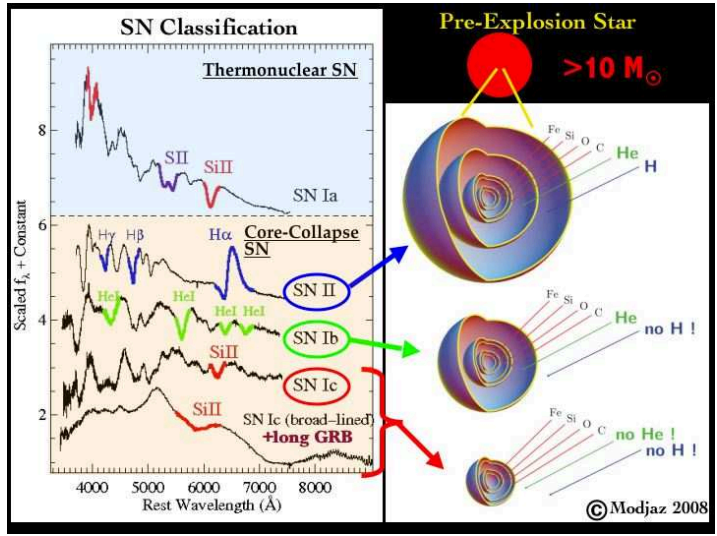}

\caption{Mapping between different types of core-collapse SNe ({\it left}) and their corresponding progenitor stars ({\it right}). {\it Left}: Representative observed spectra of different types of SNe. Broad-lined SN Ic are the only type of SNe seen in conjunction with GRBs. Not shown are some of the other H-rich SN members (SNe IIn and very luminous SNe). {\it Right}: Schematic drawing of massive ($\ge$ 8$-10$ \sm) stars before explosion, with different amounts of intact outer layers, showing the "onion-structure" of different layers of elements that result from successive stages of nuclear fusion during the massive stars' lifetimes (except for H). This figure is can be found at 
http://cosmo.nyu.edu/~mmodjaz/research.html.}

 \label{fig1}
\end{figure*}

Stripped supernovae (SNe) and long-duration Gamma-Ray Bursts (long GRBs) are nature's most powerful explosions from massive stars. They energize and enrich the interstellar medium, and, like beacons, they are visible over large cosmological distances.
However, the mass and metallicity range of their
progenitors is not known, nor the detailed physics of the explosion (see reviews by \citealt{woosley06_rev} and \citealt{smartt09_rev}). Stripped-envelope SNe (i.e, SNe of Types IIb, Ib, and Ic , e.g., \citealt{filippenko97_review}) are core-collapse events whose massive progenitors have been stripped of progressively larger
amounts of their outermost H and He envelopes (Fig.~\ref{fig1}). In particular, broad-lined SNe~Ic (SNe~Ic-bl) are SNe Ic whose line widths approach 20,000$-$30,000 \kms\ around maximum light  (see below) and whose optical spectra show no trace of H and He. 

For the last 15 years, the exciting connection between long GRBs and
SNe~Ic-bl, the only type of SNe observed accompanying long GRBs (for reviews, see \citealt{woosley06_rev,hjorth11,modjaz11_rev}), and the existence of many more SNe~Ic-bl {\it without} GRBs raises the question of what distinguishes SN-GRB progenitors from those of 
ordinary SNe~Ic-bl without GRBs. Viewing angle effects are probably not the reason why those SNe Ic-bl did not show an accompanied GRBs  \citep{soderberg06_radioobs} and based the same radio upper-limits, only $\sim$ 1\% of SNe Ib/c appear to be accompanied by GRBs \citep{soderberg10_09bb}. One promising line of
attack is to investigate what sets apart SNe Ib/c \textbf{with} GRBs from
those \textbf{without} GRBs to elucidate the conditions and progenitors of these two types of explosions. While of course there are numerous possible avenues (for a recent review see e.g., \citealt{modjaz11_rev}), I will here adopt a two-thronged approach, given the short amount of time:  First, I focus on comparing the optical spectra of SNe Ib/c with and without GRBs, since early-time optical spectra are used for identifying the spectral features of different explosions and probe the bulk of the ejected stellar material, in particular the outermost layers. Secondly, I present a meta-analysis of published measured metallicities at the explosion site of SNe Ib/c with and without GRBs. Metallicity is expected to strongly impact the lives and deaths of stars due to the metallicity dependence of mass loss (e.g., \citealt{vink05}) and its subsequent link to rotation and angular momentum content of the stellar core.  The main thrust of my talk is that now a number of different groups, including ours, have contributed to gathering large data-sets, whose analysis can to lead to robust statistical conclusions  and interesting insights into different populations of SNe with and without GRBs. 

\section{Optical Spectra and Expansion Velocities of SNe Ib/c with and without GRBs}

While the observational hallmark of a SN Ic-bl is, by the definition, its high expansion velocity (which, when  modeled in combination with light curves, yields high energies, sometimes above $10^{52}$ erg, i.e. 10 times more than the canonical CCSN, and thus motivated some to call them "HyperNovae"), there are debates within the community whether such SNe Ic-bl can be robustly distinguished from "normal" SNe Ic and whether there are systematic differences between SNe Ic-bl with and without GRBs. Prior work involving synthetic models based on Monte Carlo radiative transfer codes (\citealt{mazzali09} and reference therein), while important, has included only a few normal SNe Ic and a few SN Ic-bl without GRBs, thus not yet providing a large sample. 

Here we are using the spectra from the CfA sample of Stripped SNe (Modjaz et al. in prep), as well as those from the literature (see references in Modjaz et al. in prep) to compare the absorption velocities as traced by Fe II $\lambda$5169 of different kinds of SNe Ic. Spectral synthesis studies have shown that this and other Fe lines are good tracers of the photospheric velocity, since they do not saturate \citep{branch02}. With the largest sample of spectra to date, we find that SN Ic-bl \textbf{with} GRBs have the \textbf{highest} absorption velocities (25,000$-$35,000 \kms~at maximum V-light), while SNe Ic-bl \textbf{without} GRBs have \textbf{lower} velocities (between 15,000$-$25,000 \kms~at maximum V-light),  and normal SN Ic have the lowest absorption velocities (8,000$-$15,000 \kms ). Of course, we caution that because of sever blending, specifically in SNe Ic-bl, the Fe II $\lambda$5169 line could be blended with other nearby lines in a manner such that it may compromise the velocity measurements. However other, more isolated, lines (e.g. Si II) also indicate high velocities for SN Ic-bl. 
\section{Measured Metallicities at the Explosion Sites of SNe Ib/c with and without GRBs}

\begin{figure}[!ht]
\begin{center}
 \includegraphics[width=3.6in,angle=90]{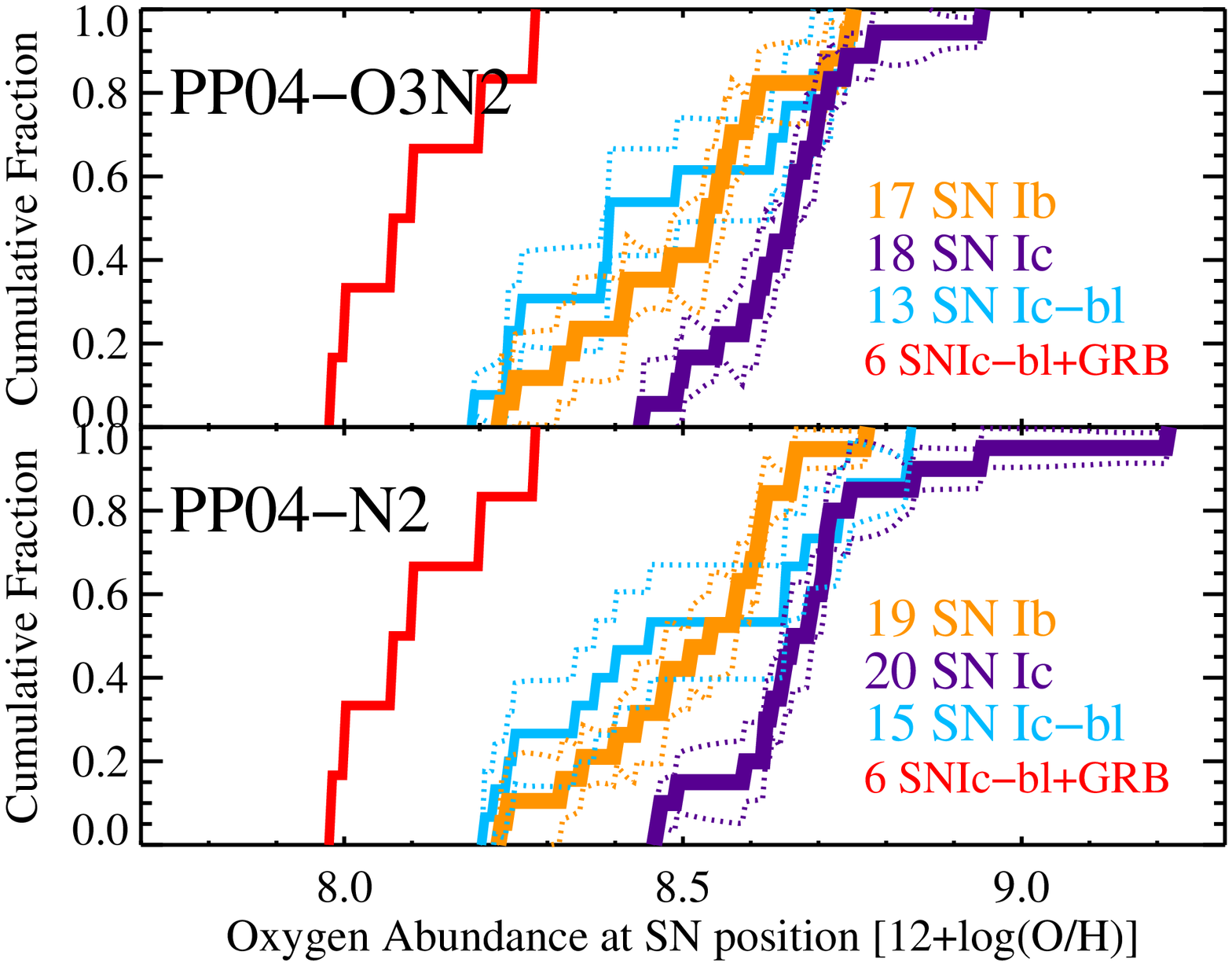} 
 \caption{Cumulative fraction (solid lines) of measured oxygen abundances at
the SN position of different types of SNe Ib/c with and without GRBs and their confidence bands (dotted lines), based on the meta-analysis of the SN samples in \citet{modjaz08_Z,anderson10,modjaz11,leloudas11} (and references therein) in the two scales of \citet{pettini04} that all samples had in common. The ordinate indicates the fraction of the SN population with metallicities less than the abscissa value. The confidence bands are shown around each cumulative trend, which we computed via bootstrap with 10,000 realizations based on the metallicity measurements and their associated reported uncertainties. SNe Ic (the demise of the most heavily stripped stars that lost much, if not all, of both their H and He layers) are systematically in more metal-rich environments than
SNe Ib (SNe arising from less stripped stars that retained their He layer). SN Ic-bl with GRBs are still at consistently lower oxygen abundances than SN Ic-bl without GRBs. The radio-loud SN Ic-bl~2009bb \citep{soderberg10_09bb} is not included in this plot, since no GRB was detected, but would add one data point at 12+log(O/H)$_{\mathrm{PP04-O3N2}}$=8.9 (Levesque et al. 2010a).  }
   \label{oxyfig}
\end{center}
\end{figure}

Since direct SN Ib/c progenitor detection attempts via deep pre-explosion images have not been successful \citep{smartt09_rev} and are impossible for GRBs, we employ a complimentary approach: we study the host galaxy environments in order to discern
any systematic trends as a function explosion type that may characterize their stellar progenitors. Specifically, massive stars at different metallicities are expected to live and die differently, due to the metallicity dependence of mass loss and its subsequent link to rotation and angular momentum content of the stellar core (e.g., \citealt{crowther07}).  Since the early work on GRB host metallicities and their comparison with SDSS galaxies as well as with SN galaxies, the field of environmental  metallicity studies has experienced a tremendous growth (see discussions in e.g., Levesque et al. 2010a, \citealt{modjaz11_rev} and references therein, as well as contributions in this volume).

Here, we outline the recipe for state-of-the-art metallicity analysis, specifically of the oxygen abundance from nebular HII region emission lines, first formalized in \citet{modjaz08_Z}: 1) In order to probe the natal oxygen abundance, obtain spectra at the position of the SN or GRB (because of metallicity gradients in spiral galaxies); 2) Include only SNe with secure SN ID (i.e., ideally from multi-epoch SN spectra to monitor for any potential classification changes) and also (only) from untargetted surveys in order to mitigate any selection effects; 3) Employ spectrographs with a large wavelength range in order to observe emission lines (from [OII] $\lambda$3727 to H$\alpha$ and [NII] $\lambda$6584) as to compute abundances in different diagnostics, since there are systematic differences between different diagnostics \citep{kewley08} ; 4) Remove stellar absorption in spectra when necessary, and 5) Obtain good handle on uncertainty budget and propagate line flux and reddening uncertainties into abundance measurement errors via Monte Carlo simulations.

While recently different groups have arrived at different conclusions about whether there is a statistically significant trend of metallicity with Stripped SN subtype (\citealt{anderson10,modjaz11,leloudas11}, see also proceedings by Anderson, Leloudas in this volume), not all measured metallicities reported in the Anderson and Leloudas et al samples are at the position of the SNe. Thus, we conducted a meta-analysis of all samples \citep{modjaz08_Z,anderson10,modjaz11,leloudas11} with the best-possible quality-control and following the above state-of-the-art recipe, now that we have larger samples to draw from: we only included oxygen abundance measurements at the exact SN explosion sites (within the slit) of SNe with solid IDs and also from untargeted surveys, to have the best handle possible on the natal metallicity estimates of SNe with well-determined SN types over a large metallicity baseline, the ultimate goal of the study. 
Figure~\ref{oxyfig} shows the result of our metal-analysis, namely the cumulative distributions of local metallicities for different types of stripped CCSNe (SNe Ib, Ic, Ic-bl without GRBs, SNe Ic-bl with GRBs) from the combined samples. We find that with a combined and large sample size, the sites of SNe Ic do indeed have higher oxygen abundances than SNe Ib, and with a higher statistical significance than in the individual samples. There is only a 0.1\% (2\%) probability in the PP04-N2 (PP04-O3N2) scale that the oxygen abundances of the 19 (17) SNe Ib and of 20 (18) SNe Ic are drawn from the same parent population, which are different on average by $\sim$0.2 dex.
Here we have taken advantage of the power of statistics by combining the hard work of three different groups.  


In addition, SN Ic-bl with GRBs still prefer, on average, more metal-poor environments than those without GRBs (see Fig. 2 in \citealt{modjaz11}), with the GRB metallicity-luminosity relation offset to lower metallicities, but without a cut-off metallicity above which GRB production would be suppressed (Levesque et al. 2010b). Since the host galaxies of both samples span similar range in galaxy luminosity (i.e,. even to luminous GRB host galaxies of $M_B=-21$ mag ), dust  effects are most likely not the reason for the offset to low metallicity. However, while these results are intriguing, the next  step is to conduct a thorough
and extensive host galaxy study with a large single-survey,
untargeted, spectroscopically classified, and homogeneous
collection of stripped SNe, something we are currently undertaking
with the Palomar Transient Factory (PTF).

M.M. acknowledges current support from the NYU ADVANCE Women-in-Science Travel Grant funded by the NSF
ADVANCE-PAID award Number HRD-0820202 and prior support from Hubble
Fellowship grant HST-HF-51277.01-A.

\begin{thebibliography}{17}
\expandafter\ifx\csname natexlab\endcsname\relax\def\natexlab#1{#1}\fi

\bibitem[{{Anderson} {et~al.}(2010){Anderson}, {Covarrubias}, {James}, {Hamuy},
  \& {Habergham}}]{anderson10}
{Anderson}, J.~P. et al.  2010, \mnras, 407, 2660

\bibitem[{{Branch} {et~al.}(2002){Branch}, {Benetti}, {Kasen}, {Baron},
  {Jeffery}, {Hatano}, {Stathakis}, {Filippenko}, {Matheson}, {Pastorello},
  {Altavilla}, {Cappellaro}, {Rizzi}, {Turatto}, {Li}, {Leonard}, \&
  {Shields}}]{branch02}
{Branch}, D., et al. 2002, \apj, 566, 1005

\bibitem[{{Crowther}(2007)}]{crowther07}
{Crowther}, P.~A. 2007, \araa, 45, 177

\bibitem[{{Filippenko}(1997)}]{filippenko97_review}
{Filippenko}, A.~V. 1997, \araa, 35, 309

\bibitem[{{Hjorth} \& {Bloom}(2011)}]{hjorth11}
{Hjorth}, J. \& {Bloom}, J.~S. 2011, {Chapter 9 in "Gamma-Ray Bursts" (arXiv:1104.2274)}

\bibitem[Kewley \& Ellison(2008)]{kewley08} Kewley, L.~J., \& Ellison, S.~L.\ 2008, \apj, 681, 1183 


\bibitem[{{Leloudas} {et~al.}(2011){Leloudas}, {Gallazzi}, {Sollerman},
  {Stritzinger}, {Fynbo}, {Hjorth}, {Malesani}, {Micha{\l}owski},
  {Milvang-Jensen}, \& {Smith}}]{leloudas11}
{Leloudas}, G., et al. 2011, \aap, 530, A95


\bibitem[Levesque et al.(2010)]{levesque10_09bb} Levesque, E.~M., et al.\ 2010a, \apjl, 709, L26 

\bibitem[{{Levesque} {et~al.}(2010){Levesque}, {Berger}, {Kewley}, \&
  {Bagley}}]{levesque10_grbhosts}
{Levesque}, E.~M., {Berger}, E., {Kewley}, L.~J., \& {Bagley}, M.~M. 2010b, \aj,
  139, 694




\bibitem[{{Mazzali} {et~al.}(2009){Mazzali}, {Deng}, {Hamuy}, \&
  {Nomoto}}]{mazzali09}
{Mazzali}, P.~A., {Deng}, J., {Hamuy}, M., \& {Nomoto}, K. 2009, \apj, 703,
  1624

\bibitem[{{Modjaz} {et~al.}(2008){Modjaz}, {Kewley}, {Kirshner}, {Stanek},
  {Challis}, {Garnavich}, {Greene}, {Kelly}, \& {Prieto}}]{modjaz08_Z}
{Modjaz}, M., et al. 2008,  \aj, 135, 1136


\bibitem[{{Modjaz} {et~al.}(2011){Modjaz}, {Kewley}, {Bloom}, {Filippenko},
  {Perley}, \& {Silverman}}]{modjaz11}
{Modjaz}, M., et al 2011, \apjl, 731, L4

\bibitem[{{Modjaz}(2011)}]{modjaz11_rev}
{Modjaz}, M. 2011, AN, 332, 434


\bibitem[{{Pettini} \& {Pagel}(2004)}]{pettini04}
{Pettini}, M. \& {Pagel}, B.~E.~J. 2004, \mnras, 348, L59

\bibitem[{{Smartt}(2009)}]{smartt09_rev}
{Smartt}, S.~J. 2009, \araa, 47, 63

\bibitem[Soderberg et al.(2006)]{soderberg06_radioobs} Soderberg, A.~M., et al. 2006, \apj, 638, 930 


\bibitem[{{Soderberg} {et~al.}(2010){Soderberg}}]{soderberg10_09bb}
{Soderberg}, A.~M., et al. 2010, \nat, 463, 513

\bibitem[{{Vink} \& {de Koter}(2005)}]{vink05}
{Vink}, J.~S. \& {de Koter}, A. 2005, \aap, 442, 587

\bibitem[{{Woosley} \& {Bloom}(2006)}]{woosley06_rev}
{Woosley}, S.~E. \& {Bloom}, J.~S. 2006, \araa, 44, 507

\end{thebibliography}


\begin{discussion}

\discuss{Anderson (Q)}{Is there is a bias in PTF for studying SNe in dim hosts? As a community we have to be careful not to over-emphasize them.}

\discuss{Modjaz(A)}{As we had discussed in Sydney, PTF has the current strategy of making sure to obtain spectroscopic classification of SNe in low-luminosity hosts, so that we have a complete view of what kinds of SNe erupt in the neglected bin of low-luminosity galaxies. However, this selection should not skew the results of demographic studies -- the SN IDs are obtained in those low-luminosity galaxies independent of the SN type. But I agree that this strategy would not lend itself for computing SN rates if there is no well-defined selection function.  }

\discuss{Ryan Chornock}{Do you see signs of Wolf-Rayet stars in the spectra of the HII regions at the explosion sites?}
\discuss{Modjaz(A)}{For some of them, yes, and a current student of mine, David Fierroz, is analyzing the data $-$ the published spectra, as well as other ones we obtained from Keck.}


\end{discussion}

\end{document}